\documentclass[twoside,12pt]{report}
\usepackage{graphicx}
\usepackage{amssymb, amsmath, amsxtra}
\usepackage{epsfig,subfigure}

\usepackage{latexsym}
\usepackage{bm}

\begin{document}
 \begin{center}
 {\Large \bf Self-adjointness of a generalized Camassa-Holm equation}\\[2ex]
 N.H.~Ibragimov, R.S.~Khamitova\\
 \textit{Department of Mathematics and Science,\\ Blekinge Institute of
Technology,\\ SE-371 79 Karlskrona, Sweden}\\
 and\\
 A. Valenti\\
 \textit{Department of Mathematics and Computer Sciences,\\
University of Catania, I 95125 Catania, Italy}

 \end{center}

 \begin{small}
 \noindent
{\bf Abstract}\\
It is well known that the Camassa-Holm equation possesses numerous
remarkable properties characteristic for KdV type equations. In this
paper we show that it shares one more property with  the KdV
equation. Namely, it is shown in \cite{ibra06,ibr07a} that the KdV
and the modified KdV equations are self-adjoint. Starting from the
generalization \cite{CMP97} of the Camassa-Holm equation
\cite{ch93}, we prove that the Camassa-Holm equation is
self-adjoint. This property is important, e.g. for constructing
conservation laws associated with symmetries of the equation in
question. Accordingly, we construct conservation laws for the
generalized Camassa-Holm equation using its symmetries.\\


 \noindent
 \textit{AMS classification numbers}:
 35C99, 35G20, 35L65, 22E70, 35Q35, 35Q51
 \end{small}

\section{Introduction}

 The Camassa-Holm equation
 \begin{equation}
 \label{ch:eq1}
 F\equiv u_t-u_{txx}-u\,u_{xxx}-2\,u_xu_{xx}+3\,u\,u_x+\kappa\,u_x=0
 \end{equation}
 has appeared in \cite{ch93}, \cite{ch94} as a shallow water wave equation. Here $u(t,x)$  is the fluid velocity in the $x$ direction and
 $\kappa$ is an arbitrary constant. Eq. (\ref{ch:eq1}) was studied also by Fokas \cite{Fok95} and Fuchsstainer \cite{Fuc96}.
\par
 Clarkson, Mansfield and Priestley \cite{CMP97} studied the third-order nonlinear equation of the form
  \begin{equation}
 \label{ch:eq1a}
 F\equiv u_t-\epsilon\,u_{txx}-u\,u_{xxx}-\beta\,u_xu_{xx}-\alpha\,u\,u_x+\kappa\,u_x=0
 \end{equation}
 with arbitrary parameters $\epsilon,\,\alpha,\,\beta .$  It contains not only the Camassa-Holm equation (\ref{ch:eq1})
 as a particular case, but also other interesting nonlinear equations such as:
 \begin{itemize}
 \item
 the Fornberg-Whitham equation \cite{FW}
$$
 u_t-\,u_{txx}-u\,u_{xxx}-3\,u_xu_{xx}+u\,u_x+u_x=0,
$$
  \item
  the Rosenau-Hyman equation \cite{RH}
$$
 u_t-u\,u_{xxx}-3\,u_xu_{xx}-u\,u_x=0.
$$
\end{itemize}
\par
 Eq. (\ref{ch:eq1}), as any evolution type equation, does not have a usual Lagrangian\footnote{By this we mean that
 there is no function $L(x, t, u, u_x, u_t, u_{xx}, u_{tx}, \ldots)$ such that Eq. (\ref{ch:eq1}) is identical with
 the Euler-Lagrange equation
 $$
 \frac{\delta L}{\delta u} \equiv \frac{\partial L}{\partial u}- D_x\frac{\partial L}{\partial u_x}
 - D_t\frac{\partial L}{\partial u_t} + D_x^2\frac{\partial L}{\partial u_{xx}}
 + D_tD_x\frac{\partial L}{\partial u_{tx}} + \cdots = 0.
 $$
However it is shown in \cite{holm-mar98} that the CH equation can
be written as the Euler-Poincar\'{e} equation.
 It also admits two canonical Hamiltonian representations,
 namely in terms of the \textit{Clebsch variables} and the so-called \textit{peakon
 variables}, obtained by using different momentum maps.}.
 Therefore the classical Noether theorem cannot be employed
 for constructing conservation laws using symmetries of Eq. (\ref{ch:eq1}). On the other hand, a new procedure was developed in \cite{ibr07a}
 for constructing conservations laws associated with symmetries. The new procedure
 allows one to construct a conservation law using any (Lie point, Lie-B\"{a}cklund, nonlocal, etc.) symmetry of any differential equation.
 However, the resulting conservation laws involve, in general, not only the solutions of the original equation, but also so-called
 \textit{nonlocal variables}, namely solutions of the adjoint equation. The nonlocal variables can be eliminated if the equation
 under consideration is quasi self-adjoint (or, in particular, self-adjoint) in the sense defined in \cite{ibr07d}.
 Therefore the quasi self-adjointness  is important for constructing conservation laws.
 Accordingly, we start our paper with investigating the quasi self-adjointness of the generalized Camassa-Holm equation (\ref{ch:eq1a}).
 Our construction require the concepts of the \textit{formal Lagrangian} and the \textit{adjoint equation} for Eq. (\ref{ch:eq1a}).

 \section{Formal Lagrangian and adjoint equation}
 \label{ch:s1}

According to the procedure suggested in \cite{ibr07a},  we introduce
the \textit{formal Lagrangian}
 \begin{equation} \label{ch:eq41}
 {\cal{L}}\equiv vF=v\left[u_t-\epsilon\,u_{txx}-uu_{xxx}-\beta\,u_xu_{xx}-\alpha\,uu_x+\kappa\,u_x\right]
\end{equation}
 and define the
adjoint equation $F^*=0$ by
\begin{equation}
 \label{ch:eq42}
F^*\equiv\frac{\delta {\cal{L}}}{\delta u}=0.
\end{equation}
Here $v=v(t,x)$ is a new dependent variable,
\[\frac{\delta}{\delta u}= \frac{\partial}{\partial u}- D_i\frac{\partial}{\partial u_i}
+D_iD_j\frac{\partial}{\partial
u_{ij}}-D_iD_jD_k\frac{\partial}{\partial u_{ijk}}+...\qquad
(i,j,k=1,2),
\]
is the variational derivative and
$$D_i= \frac{\partial}{\partial x^i}+ u_i\frac{\partial}{\partial u}+ v_i\frac{\partial}{\partial v}
+u_{ij}\frac{\partial}{\partial u_j}+v_{ij}\frac{\partial}{\partial
v_j} +u_{ijk}\frac{\partial}{\partial
u_{jk}}+v_{ijk}\frac{\partial}{\partial v_{jk}}+...$$ is the
operator of total differentiation with respect to $x^i$
($x^1=t,\,x^2=x).$ The usual convention of summation over repeated
indices is used.

 We have
\begin{equation}
F^*\equiv\frac{\partial\cal{L}}{\partial u}- D_t\frac{\partial
\cal{L}}{\partial u_t}-D_x\frac{\partial \cal{L}}{\partial u_x}
+D_x^2\frac{\partial \cal{L}}{\partial u_{xx}}-D_x^3\frac{\partial
\cal{L}}{\partial u_{xxx}} -D_tD_x^2\frac{\partial \cal{L}}{\partial
u_{txx}} =0
\end{equation}
and, taking into account the relation (\ref{ch:eq41}), we obtain
\begin{eqnarray}
F^*=&-& vu_{xxx}-\alpha\,vu_x-D_t(v)-D_x(-\beta\,vu_{xx}-\alpha\,vu+\kappa\,v) \nonumber \\[1ex]
&+& D_x^2(-\beta\,vu_x)-D_x^3(-vu)+\epsilon\,D_tD_x^2(v). \nonumber
\end{eqnarray}
Performing here the differentiations, we arrive at the following
adjoint Eq. (\ref{ch:eq42}):
\begin{equation}
 \label{ch:eq43}
F^*\equiv
-v_t+\epsilon\,v_{txx}+uv_{xxx}+(3-\beta)(u_xv_{xx}-v_xu_{xx})+\alpha\,uv_x-\kappa\,v_x=0.
\end{equation}
 \textbf{Definition 1.}
An equation $F=0$  is said to be \textit{quasi self-adjoint}
\cite{ibr07d} if there exists a function
\begin{equation}
 \label{ch:eq44s}
v=\varphi(u), \quad \varphi'(u)\neq 0,
\end{equation}
such that
\begin{equation}
 \label{ch:eq44}
F^*\mid_{v=\varphi(u)} = \lambda\,F
\end{equation}
with an undetermined coefficient $\lambda$. If in (\ref{ch:eq44s})
$\varphi (u) = u,$ we say that the equation $F=0$ is
\textit{self-adjoint.}

Taking into account the expression (\ref{ch:eq43}) of $F^*$ and
using Eq. (\ref{ch:eq44s}) together with its consequences
\[
v_t=\varphi'u_t,\quad v_x=\varphi'u_x,
\]
\[
v_{tx}=\varphi'u_{tx}+\varphi''u_tu_x,\quad
v_{xx}=\varphi'u_{xx}+\varphi''u_x^2,
\]
\[
v_{txx}=\varphi'u_{txx}+2\varphi''u_xu_{tx}+\varphi''u_tu_{xx}+\varphi'''u_tu_x^2,
\]
\[
v_{xxx}=\varphi'u_{xxx}+3\varphi''u_xu_{xx}+\varphi'''u_x^3,
\]
we rewrite Eq. (\ref{ch:eq44}) in the following form:
\begin{eqnarray}
&& -\varphi' u_{t} + \alpha\,\varphi' u\,u_x -\kappa\,\varphi' u_x + [\varphi''' u-(\beta-3)\,\varphi'')]\,u_x^3 + \varphi' u\,u_{xxx} \nonumber \\
&& +\,\epsilon\,\varphi' u_{txx} +\,[3\,\varphi''
u-2\,(\beta-3)\,\varphi')]\,u_x\,u_{xx}
+ 2\,\epsilon\,\varphi'' u_x\,u_{tx} + \epsilon\,\varphi''' u_{t}\,u_x^2   \nonumber \\
&& +\,\epsilon\,\varphi'' u_t\,u_{xx} =
\lambda\,(u_t-\epsilon\,u_{txx}-uu_{xxx}-\alpha\,uu_x-\beta\,u_xu_{xx}+\kappa\,u_x)
 \label{ch:eq45}.
\end{eqnarray}
Eq. (\ref{ch:eq45}) should be satisfied identically in all variables
$u_t,\,u_x,\,u_{xx},\,\ldots .$ Comparing the coefficients of $u_t$
in both sides of Eq. (\ref{ch:eq45}), we obtain $\lambda=-\varphi'$.
Then we equate the coefficients of $u_xu_{tx}$ and get:
\begin{equation}
 \label{ch:eq46}
\epsilon\,\varphi''=0.
\end{equation}
According to Eq. (\ref{ch:eq46}), the procedure splits into two
cases:
\begin{eqnarray}
&& \epsilon=0, \label{ch:eq47} \\
&& \epsilon\neq0,\quad \varphi''=0. \label{ch:eq48}
\end{eqnarray}
\par
In the case (\ref{ch:eq47}) we compare the coefficients of
$u_x\,u_{xx}$ and arrive at the equation
\begin{equation}
 \label{ch:eq49}
\varphi''\,u + (\beta-2)\,\varphi'=0.
\end{equation}
Integrating Eq. (\ref{ch:eq49}) we obtain
\begin{eqnarray}
 \label{ch:eq49a}
\varphi(u)=\left\{
\begin{array}{lll}
a+b\,u^{\beta-1}, \quad \beta\neq1,\\
a+b\,\ln u, \quad \beta=1,
\end{array}
\right.
\end{eqnarray}
where $a$ and $b$ are arbitrary constants. The coefficients of the
other terms in both sides of Eq. (\ref{ch:eq45}) are equal due to
Eq. (\ref{ch:eq49}). Hence we have proved that the equation
\begin{equation}
\label{ch:eq410}
u_t-u\,u_{xxx}-\beta\,u_xu_{xx}-\alpha\,u\,u_x+\kappa\,u_x=0
\end{equation}
is quasi self-adjoint and that the substitution (\ref{ch:eq44s}) has
the form
\begin{eqnarray}
\label{ch:eq411} v = \left\{
\begin{array}{lll}
a + b\,u^{\beta-1}, \quad \beta\neq1,\\
a + b\,\ln u, \quad \beta=1.
\end{array}
\right.
\end{eqnarray}
\par
Now we consider the case (\ref{ch:eq48}). In this case the
comparison of the coefficients for $u_xu_{xx}$ yields that
$\beta=2$. The coefficients of the other terms in both sides of Eq.
(\ref{ch:eq45}) are equal. Thus we have proved that the equation
\begin{equation}
\label{ch:eq412}
u_t-\epsilon\,u_{txx}-u\,u_{xxx}-2\,u_xu_{xx}-\alpha\,u\,u_x+\kappa\,u_x=0
\end{equation}
is quasi self-adjoint for any parameters
$\epsilon,\,\alpha,\,\kappa$ and that the substitution
(\ref{ch:eq44s}) has the following form:
\begin{equation}
\label{ch:eq413} v = a + b\,u.
\end{equation}
We can take, in particular, $a=0$ and $b=1$. Hence Eq.
(\ref{ch:eq412}) coincides with its adjoint equation after the
substitution $v=u$. According to \cite{ibra06}, it means that the
\textit{generalized Camassa-Holm equation} (\ref{ch:eq412}) is
self-adjoint.
\par
In conclusion of this section we note that  the Fornberg-Whitham
equation
\begin{equation}
\label{ch:eq1b} u_t-\,u_{txx}-u\,u_{xxx}-3\,u_xu_{xx}+u\,u_x+u_x=0
\end{equation}
is not quasi self-adjoint in the sense of Definition 1.

\section{Conservation laws}
\label{ch:s3}
\subsection{General form}
\label{ch:s31} We rewrite the formal Lagrangian (\ref{ch:eq41}) in
the symmetric form
 \begin{equation} \label{ch:eq51}
 {\cal{L}}=v\left[u_t-\frac {\epsilon} {3} (u_{txx}+u_{xtx}+u_{xxt})-u\,u_{xxx}-\alpha\,u_xu_{xx}-\beta\,u\,u_x+\kappa\,u_x\right].
\end{equation}
Eq. (\ref{ch:eq1a}) is said to have a \textit{nonlocal conservation
law }if there exits a vector \mbox{$\textbf{C}=(C^1, C^2)$}
satisfying the equation
\begin{equation}
 \label{ch:eq52}
D_t(C^1)+D_x(C^2)=0
\end{equation}
on any solution of the system (\ref{ch:eq1a}), (\ref{ch:eq43}). Eq.
(\ref{ch:eq1a}) has a \textit{local conservation law} if
(\ref{ch:eq52}) is satisfied on any solution of Eq. (\ref{ch:eq1a}).
\par
The conserved vector corresponding to an operator
\begin{equation}
 \label{ch:eq53}
X=\xi^1(t,x,u)\frac{\partial }{\partial
t}+\xi^2(t,x,u)\frac{\partial }{\partial
x}+\eta(t,x,u)\frac{\partial }{\partial u}
\end{equation}
 admitted by Eq. (\ref{ch:eq1a}) is obtained by the following formula \cite{ibr07a}:
\begin{eqnarray}
&& C^i=\xi^{i}{\cal L}+W\left[\frac{\partial {\cal L}}{\partial u_i
}-D_j\left(\frac{\partial {\cal L}}{\partial u_{ij} }\right)+
D_jD_k\left(\frac{\partial {\cal L}}{\partial u_{ijk} }\right)\right] \nonumber \\[2ex]
&& \qquad \,+\,D_j(W)\left[\frac{\partial {\cal L}}{\partial u_{ij}
}-D_k\left(\frac{\partial {\cal L}}{\partial u_{ijk}
}\right)\right]+ D_jD_k(W)\frac{\partial {\cal L}}{\partial u_{ijk}
},  \label{ch:eq54}
\end{eqnarray}
where  $i,j,k=1,2$ and
\[
 W=\eta -\xi^{i}u_i.
 \]
\par
We will construct the conserved vectors (\ref{ch:eq54}) using the
Lie point symmetries (\ref{ch:eq53}) of \mbox{Eq. (\ref{ch:eq1a})}
found in \cite{CMP97}.

\subsection{Rosenau-Hyman equation}
\label{ch:s32}

Letting in Eq. (\ref{ch:eq410}) $\alpha=1,\, \beta=3,\,\kappa=0$, we
obtain the Rosenau-Hyman equation
\begin{equation}
 \label{ch:eq1c}
 u_t-u\,u_{xxx}-3\,u_xu_{xx}-u\,u_x=0.
  \end{equation}
  In this case, the formal Lagrangian (\ref{ch:eq51}) and the substitution (\ref{ch:eq49a}) assume  the  forms
   \begin{equation} \label{ch:eq55}
 {\cal{L}}=v\left[ u_t-uu_{xxx}-3\,u_xu_{xx}-uu_x\right]
\end{equation}
and
   \begin{equation} \label{ch:eq56}
v=a+bu^2,
\end{equation}
respectively. We construct the conservation law associated with the
scaling symmetry
\begin{equation}
 \label{ch:eq57}
X=u\,\frac{\partial }{\partial u}-t\,\frac{\partial }{\partial
t}\,\cdot
\end{equation}
For this symmetry we have $W=u+t\,u_t$. Writing the quantities
(\ref{ch:eq54}) without the term $\xi^{i}{\cal L}$  since the
Lagrangian ${\cal L}$ is equal to zero on solutions of Eq.
(\ref{ch:eq1c}) and taking into account the structure of the formal
Lagrangian (\ref{ch:eq55}),  we obtain
\begin{eqnarray}
&& C^1=W\,\frac{\partial {\cal L}}{\partial u_t},  \label{ch:eq57a} \\[2ex]
&& C^2=W\left[\frac{\partial {\cal L}}{\partial u_x}
-D_x\left(\frac{\partial {\cal L}}{\partial u_{xx} }\right)
+D_x^2\left(\frac{\partial {\cal L}}{\partial u_{xxx} }\right)\right] \nonumber \\[2ex]
&& \qquad \,+\,D_x(W)\left[\frac{\partial {\cal L}}{\partial u_{xx}
} - D_x\left(\frac{\partial {\cal L}}{\partial u_{xxx}}\right)
\right] +D_x^2(W)\frac{\partial {\cal L}}{\partial u_{xxx}}.
\label{ch:eq57b}
\end{eqnarray}
Substituting in (\ref{ch:eq57a}) and (\ref{ch:eq57b}) the expression
(\ref{ch:eq55}) for ${\cal{L}}$, we get
\begin{eqnarray}
&& C^1=vW,  \label{ch:eq57c} \\[1ex]
&& C^2=\left(-u\,v+u_xv_x-v\,u_{xx}-u\,v_{xx}\right)W \nonumber \\[1ex]
&& \qquad +\left(u\,v_x-2\,v\,u_x\right)D_x(W)-u\,v\,D_x^2(W).
\label{ch:eq57d}
\end{eqnarray}
Now we substitute in Eq. (\ref{ch:eq57c}) the expression
$W=u+t\,u_t,$ eliminate  $u_t$ by using Eq. (\ref{ch:eq1c})  and
obtain:
\begin{eqnarray}
&& C^1 = u\,v+t\,v\left(u\,u_{xxx}+u\,u_x+3\,u_xu_{xx}\right)  \nonumber \\
&& \quad \,\,\,= u\,v+t\,u\,v\,u_x-\frac 3 2 \,t\, v_xu_x^2-t\,D_x\left(u\,v\right)u_{xx} \nonumber \\
&& \qquad \,+\,D_x\left(t\,u\,v\,u_{xx}+\frac 3 2
\,t\,v\,u_x^2\right). \label{ch:eq57e}
\end{eqnarray}
We can shift the last term in Eq. (\ref{ch:eq57e}) into $C^2$ by
using the identity
$$
D_t({\tilde C}^1+D_x(A)) + D_x(C^2)=D_t({\tilde C}^1) +
D_x(C^2+D_t(A))
$$
and obtain
 \begin{equation}
C^1= u\,v+t\,u\,v\,u_x-\frac 3 2 \,t\,
v_xu_x^2-t\,D_x\left(u\,v\right)u_{xx}. \label{ch:eq57f}
\end{equation}
Now we substitute in Eq. (\ref{ch:eq57f}) the expression
(\ref{ch:eq56}) for $v$, shift the terms of the form $D_x(\cdots)$
into $C^2$ and finally  arrive at the conserved vector with the
following components:

\begin{equation}
 \label{ch:eq58}
 \begin{split}
&C^1 = a\,u+b\,u^3, \\[1.5ex]
&C^2 = -a\left(\frac 1 2 \,u^2 + u_x^2 + u\,u_{xx}\right) -
b\left(\frac 3 4 \,u^4 + 3\,u^3\,u_{xx}\right).
\end{split}
\end{equation}
The vector (\ref{ch:eq58}) is a linear combination with constant
coefficients $a$ and $b$ of the following two linearly independent
conserved vectors:
\begin{equation}
 \label{ch:eq59}
C^1 = u, \,\,\qquad C^2 = -\frac 1 2 \,u^2 - u_x^2 - u\,u_{xx}
\end{equation}
and
\begin{equation}
 \label{ch:eq510}
C^1 = u^3, \qquad C^2 = -\frac 3 4 \,u^4 - 3\,u^3\,u_{xx}.
\end{equation}
\par
The conservation equation (\ref{ch:eq52}) for the vector
(\ref{ch:eq59}) coincides with Eq. (\ref{ch:eq1c}), whereas the
vector (\ref{ch:eq510}) provides a new conservation law for the
Rosenau-Hyman equation.

\subsection{Camassa-Holm equation}
\label{ch:s33} For the Camassa-Holm equation (\ref{ch:eq1}) the
formal Lagrangian (\ref{ch:eq51}) is written
 \begin{equation} \label{ch:eq511}
 {\cal{L}}=v\left[u_t-\frac {1} {3} (u_{txx}+u_{xtx} + u_{xxt})-uu_{xxx}-2\,u_xu_{xx}+3\,uu_x+\kappa\,u_x\right].
\end{equation}
We will construct the conservation law by taking the substitution
(\ref{ch:eq413}) of the particular form $v=u$. We use the following
symmetry of Eq. (\ref{ch:eq1}):
\begin{equation}
 \label{ch:eq513}
X=-2\,t\frac{\partial }{\partial t}+\kappa\,t\frac{\partial
}{\partial x}+\left(\kappa +2\,u\right)\frac{\partial }{\partial
u}\,\cdot
\end{equation}
Proceeding as in Section \ref{ch:s32}, we obtain the following
conserved vector associated with the symmetry (\ref{ch:eq513}):
\begin{equation}
 \label{ch:eq514}
 \begin{split}
&C^1 = 2(u^2 + u_x^2) + \kappa\,u, \\[1.5ex]
&C^2 = 4(u^3 - u^2\,u_{xx} - u\,u_{tx}) + \kappa\left(\frac 7 2
\,u^2 - \frac 1 2 \,u_x^2 - u\,u_{xx} - u_{tx} + \kappa\,u \right) .
\end{split}
\end{equation}
When $\kappa = 0$ in Eq. (\ref{ch:eq1}) the symmetry
(\ref{ch:eq513}) takes the form (\ref{ch:eq57}), and the conserved
vector (\ref{ch:eq514}) becomes
\begin{equation}
 \label{ch:eq512}
C^1 = u^2 + u_x^2, \qquad C^2 = 2\left(u^3 - u^2u_{xx} -
u\,u_{tx}\right).
\end{equation}
\par
It is shown in  \cite{ibra06}  that the well known infinite series
of conservation laws of the KdV equation can be obtained by
applying the formulae  (\ref{ch:eq54}) to the infinite set of
Lie-B\"{a}cklund and non-local symmetries of the KdV equation. The
similar procedure can be applied to  the Camassa-Holm and
Rosenau-Hyman equations. This is a topic for further research.

\section*{Acknowledgements}
\noindent Authors acknowledge the support of the University of
Catania through P.R.A. 2009 "{\it Metodi e Modelli Matematici per
Applicazioni di Interesse Fisico e Biologico}". N.H.I. and R.K.
want to thank the Department of Mathematics and Computer Sciences
of Catania University for the hospitality during the period in
which this work has been done. We express our gratitude to Darryl
Holm for bringing to our attention the paper \cite{holm-mar98} and
a useful discussion on canonical Hamiltonian representations of
the CH equation.

\end{document}